\documentclass[reprint,aps,prx,letterpaper]{revtex4-2}
\usepackage{amssymb}
\usepackage{amsmath}
\usepackage{graphicx}
\usepackage{footnote}
\usepackage[usenames,dvipsnames,svgnames,table]{xcolor}
\usepackage[T1]{fontenc}
\usepackage[
pdfauthor={Paul D. Nation, Hwajung Kang, Neereja Sundaresan, Jay M. Gambetta},
pdftitle={Scalable mitigation of measurement errors on quantum computers},
bookmarks=true,colorlinks=true,linkcolor=OrangeRed,urlcolor=RoyalBlue,citecolor=DarkOrchid]{hyperref}

\definecolor{lightgray}{gray}{0.95}
\let\oldtabular\tabular
\let\endoldtabular\endtabular
\renewenvironment{tabular}{\rowcolors{2}{white}{lightgray}\oldtabular}{\endoldtabular}

\begin{document}

\title{Scalable mitigation of measurement errors on quantum computers}

\author{Paul D. Nation}
\email[E-mail: ]{paul.nation@ibm.com}
\author{Hwajung Kang}
\author{Neereja Sundaresan}
\author{Jay M. Gambetta}
\affiliation{IBM Quantum, Yorktown Heights, NY 10598 USA}

\begin{abstract}
We present a method for mitigating measurement errors on quantum computing platforms that does not form the full assignment matrix, or its inverse, and works in a subspace defined by the noisy input bit-strings. This method accommodates both uncorrelated and correlated errors, and allows for computing accurate error bounds.  Additionally, we detail a matrix-free preconditioned iterative solution method that converges in $\mathcal{O}(1)$ steps that is performant and uses orders of magnitude less memory than direct factorization.  We demonstrate the validity of our method, and mitigate errors in a few seconds on numbers of qubits that would otherwise be intractable.
\end{abstract}
\date{\today}

\maketitle

\section{Introduction}\label{sec:intro}
Recently, rapid developments in the fabrication, control, and deployment of quantum computing systems has brought qubit counts to $\sim 100$, where it might be possible to show advantage over classical computation methods in one or more limited cases \cite{bravyi:2018, bravyi:2020b, maslov:2021}.  However, such breakthroughs are hampered by noise and errors that conspire to limit the effectiveness of quantum computers at tackling problems of appreciable scale.  To counteract these effects, researchers have turned to mitigation methods that approximately cancel quantum gate \cite{temme:2017, endo:2018, kandala:2019, mcardle:2019b, giurgica:2020, sun:2021, kim:2021} and measurement assignment   \cite{bravyi:2020, geller:2020, geller:2020b, hamilton:2020, ewout:2020, mac:2020, nachman:2020, hicks:2021, wang:2021} errors.  For short-depth quantum circuits that can be executed on current generation hardware, measurement errors play an out-sized role, and their correction is critical to many near-term experiments \cite{kandala:2017, gong:2019, havlicek:2019, wei:2020, mooney:2021, mooney:2021b, satzinger:2021, glick:2021}.

In the canonical situation where initialization noise is minimal, measurement errors over $N$-qubits can be treated classically and satisfy
\begin{equation}\label{eq:start}
\vec{p}_{\rm noisy} = A \vec{p}_{\rm ideal},
\end{equation}
where $\vec{p}_{\rm noisy}$ is a vector of noisy probabilities returned by the quantum system, $\vec{p}_{\rm ideal}$ is the probabilities in absence of measurement errors (but still includes e.g. gate errors), and $A$ is the $2^{N}\times 2^{N}$ complete assignment matrix (A-matrix) where element $A_{\tt row, \tt col}$ is the probability of bit-string $\tt col$ being converted to bit-string $\tt row$ by the measurement error process [see App.~(\ref{app:amat}) for examples].  While computing $A$ requires executing $2^{N}$ circuits, it is often the case that errors on multiple qubits can be well approximated using at most $\mathcal{O}(N)$ calibration circuits; the A-matrix can be approximated efficiently.

Equation~(\ref{eq:start}) has a solution $\vec{p}_{\rm ideal}$ readily found using direct LU-factorization.  However, direct methods necessarily generate quasi-probability distributions due to finite sampling that contain negative values, but still sum to one, that are incompatible with the requirement of $\vec{p}_{\rm ideal}$ being a probability vector.  Consequently, a bounded-minimization approach solving $||A\vec{p}_{\rm ideal}-\vec{p}_{\rm noisy}||^{2}_{2}$, where $\vec{p}_{\rm ideal}$ is constrained to be positive, is often used in place of a direct solution \cite{geller:2020, geller:2020b, hamilton:2020, mac:2020, Qiskit}.  Although physically appealing, the run times of these methods are orders of magnitude longer than those of direct techniques.  Alternatively, it has been shown that quasi-probabilities can be used provided that one mitigates expectation values \cite{pashayan:2015, temme:2017, bravyi:2020}.  As proven in Ref.~\cite{bravyi:2020}, these quasi-probabilities provide an unbiased estimate for the expectation value $\xi$ of an operator $O$, with a spectral radius of one, that is diagonal in the computational basis

\begin{equation}\label{eq:expval}
\xi = \sum_{i=0}^{2^{N}-1}\left[OA^{-1}\vec{p}_{\rm noisy}\right]_{i}.
\end{equation}

Near-term algorithms such as the ubiquitous Variational Quantum Eigensolver (VQE) \cite{peruzz:2014, kandala:2017} and quantum machine learning \cite{havlicek:2019, glick:2021} rely on the computation of expectation values, making the correction of measurement errors in these quantities an important step along the road to quantum advantage.

\begin{table}[b]
\begin{tabular}{l|ccccccc}
System & Avg. assignment error (\%) \\
\hline
Alibaba 11Q \cite{alibaba-hw} & 7.4 \\
Google 53Q Sycamore \cite{google-hw} & 3.2 \\
IBM 27Q Falcon\_R5.11 \cite{iqx-systems} & 1.1 \\
IONQ 11Q \cite{aws-hw} & 0.4 \\
Quantum Inspire 5Q Starmon \cite{inspire-hw} & 4.0 \\
Rigetti 32Q Aspen-9 \cite{aws-hw} & 6.1 \\
\end{tabular}
\caption{Representative error rates for cloud-accessible quantum computing systems.}
\label{table:1}
\end{table}

Current measurement mitigation techniques utilize the full $2^{N}$-dimensional probability space, and thus do not scale beyond a handful of qubits.  A truncation scheme was developed in \cite{wei:2020}, however it did so at the loss of measurement information, and still required explicit construction of the full A-matrix.  Creating a scalable mitigation strategy requires reducing the dimensionality of the linear system in Eq.~(\ref{eq:start}) without the need for computing $A$ itself.  Fortunately, present day cloud-accessible quantum computing systems have measurement error rates of a few-percent or less, see Table~\ref{table:1}, indicating it is possible to view the measurement error process as a small correction to the ideal probability distribution; measurement errors redistribute small fractions of probability from a given bit-string primarily to those that are a short Hamming distance away [e.g. see App.~(\ref{app:amat})].  To good approximation, the solution is contained within $\vec{p}_{\rm noisy}$, and we can mitigate errors in a re-normalized subspace defined by these bit-strings.  Worst case, this subspace dimension is equal to the number of times the input circuit is sampled.  For cloud-accessible quantum computers, the number of times a circuit can be sampled is limited, typically $8192$ times on IBM Quantum systems, and therefore the dimensionality of the corresponding reduced assignment matrix $\tilde A$ can be markedly smaller than the full A-matrix for $N$-qubits.  In practice $\tilde A$ is often small enough such that the solution is amenable to standard LU-factorization, returning a vector of quasi-probabilities for use in a reduced version of Eq.~(\ref{eq:expval}).

However, for situations where explicitly forming $\tilde A$ is still prohibitive due to large numbers of unique samples, it is possible to use preconditioned matrix-free iterative linear solution methods.  Such methods have also been explored in numerical solutions of large-scale steady-state density matrices \cite{nation:2015}.  In practice this gives quick convergence, typically in $\mathcal{O}(1)$ steps, and is competitive with direct solution run times while requiring orders of magnitude less memory.  Although the methods introduced here return quasi-probabilities, it is possible to find the nearest probability distribution, in terms of $L2$-norm, in linear time \cite{smolin:2012}.

In this paper, we detail this efficient mitigation method beginning with Sec.~(\ref{sec:subspace}) that motives the subspace reduction procedure, and describes how it is performed.  Section (\ref{sec:free}) shows how preconditioned matrix-free methods can be utilized for a performant and memory efficient solution technique.  In Sec.~(\ref{sec:overhead}) we show that one can obtain bounds on the variance of the computed expectation values in a similarly efficient manner with an overhead of $\mathcal{O}(1)$ in terms of additional run time.  We demonstrate our technique in Sec.~(\ref{sec:demos}), showing the validity of our method, and mitigating readout errors out to numbers of qubits that would be intractable on even the largest of supercomputers using previous methods.

\section{Subspace reduction}\label{sec:subspace}
\begin{figure*}[t]
\includegraphics[width=\textwidth]{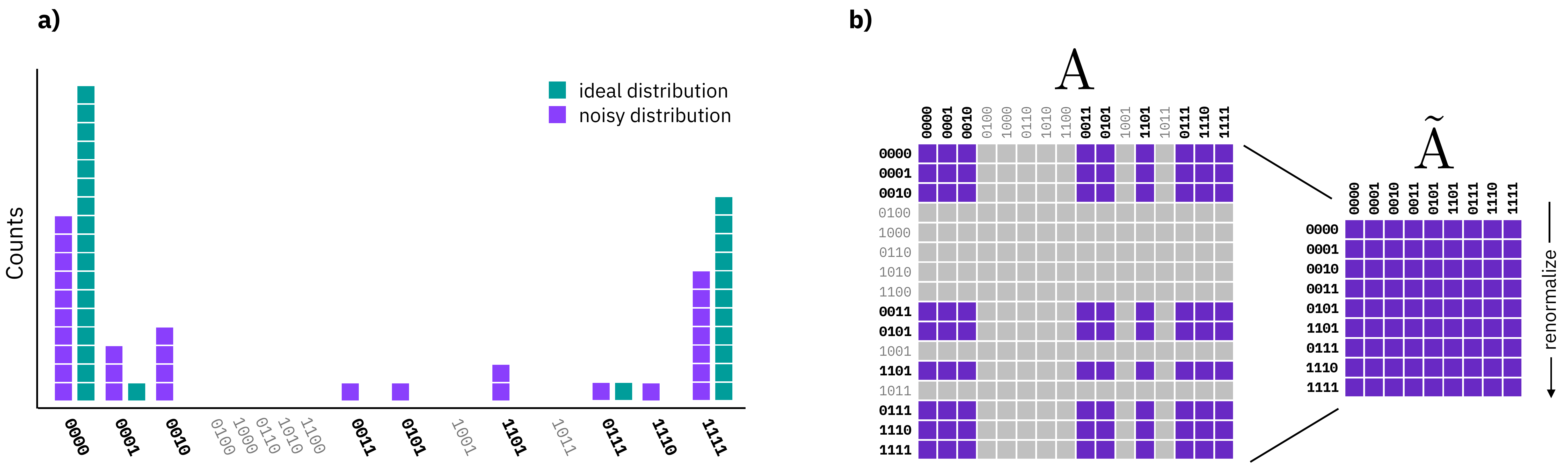}
\caption{a) Simulated discrete probability distribution for 30 samples showing the ideal distribution for a Greenberger–Horne–Zeilinger (GHZ) state subject to gate errors only as well as noisy data affected by both gate and measurement errors. Observed bit-strings are in bold.  The same random number seed was used for both simulations such that the difference between ideal and noisy distributions is solely due to measurement errors. b) Elements of the assignment matrix $A$ that are used for constructing the reduced assignment matrix $\tilde A$ that corresponds to the noisy distribution from part (a).  Columns of $\tilde A$ must be renormalized such that they again sum to one, Eq.~(\ref{eq:reduced}).}
\label{fig:1}
\end{figure*}

We aim to construct $\tilde A$ without necessarily forming the full $A$-matrix.  To this end we look to compute elements $A_{\tt row, \tt col}$, directly from the bit-string values of the input noisy counts, and a small set of calibration data matrices.  For concreteness, we study the case of tensored measurement errors as IBM Quantum systems are calibrated for high-fidelity quantum non-demolition (QND) measurements where uncorrelated errors are nominally dominant.  Other quantum hardware vendors also report the same \cite{satzinger:2021}.  The full-dimensional tensored A-matrix $A^{(T)}$ over $N$-qubits can be constructed from $N$ $2\times2$ calibration matrices: $A^{(T)}=S_{N-1}\otimes\dots S_{1}\otimes S_{0}$ where $S_{k}$ is the calibration matrix for the $k$-th qubit with the form

\begin{equation} \label{eq:1qcal}
    S_{k} = \begin{bmatrix} P^{(k)}_{0,0} & P^{(k)}_{0,1} \\
     P^{(k)}_{1,0} & P^{(k)}_{1,1}
    \end{bmatrix},
\end{equation}

where $P^{k}_{i,j}$ is the probability of the $k$-th qubit being in state $j \in \{0,1\}$ and measured in state $i \in \{0,1\}$.  Here we use the convention that qubit $0$ corresponds to the least-significant bit.  For two bit-strings: $\texttt{row},\texttt{col} \in \{0,1\}^{N}$, the matrix element $A^{(T)}_{\tt row, \tt col}$ can be computed using:

\begin{equation} \label{eq:matelem}
    A^{(T)}_{\texttt{row}, \texttt{col}} = \prod_{k=0}^{N-1}P^{(k)}_{\texttt{row}[N-1-k], \texttt{col}[N-1-k]}.
\end{equation}

Therefore it is possible to compute individual matrix elements directly from bit-string values and a number of calibration matrices that scales at most linearly with the number of qubits.  Accommodating correlated errors in our method simply amounts to finding an equivalent expression to Eq.~(\ref{eq:matelem}).  We give an example in App.~(\ref{app:correlated}), where we intentionally induce correlated errors into the measurement process.  As a corollary of grabbing elements individually, it is possible to select only those elements within a given Hamming distance, $d(\texttt{row}, \texttt{col}) \leq D$, where $D$ is the desired maximum distance.  This allows for varying the sparsity of $\tilde{A}$, and examining the effect of low-distance approximations.

$\tilde A$ is defined to be the assignment matrix over only those bit-strings observed in $\vec{p}_{\rm noisy}$.  To understand the validity of this reduction consider the simulation presented in Fig.~(\ref{fig:1}a) where we compare the distributions for $\vec{p}_{\rm ideal}$ and $\vec{p}_{\rm noisy}$.  In effect, measurement errors take blocks of probability, as defined by the finite number of samples, from $\vec{p}_{\rm ideal}$ and redistribute them amongst other bit-strings to get $\vec{p}_{\rm noisy}$.  This redistribution is predominantly to those bit-strings that are close in Hamming distance.  Importantly, as seen in Fig.~(\ref{fig:1}a), when measurement errors are small, and a sufficient number of circuit samples has been performed, it is unlikely that a given bit-string in $\vec{p}_{\rm ideal}$ is completely redistributed; the bit-strings in $\vec{p}_{\rm ideal}$ are also contained in $\vec{p}_{\rm noisy}$.  Therefore, mitigating measurement errors requires only elements $A_{\tt row, \tt col}$ that correspond to transitions between bit-strings in $\vec{p}_{\rm noisy}$.  An example is shown in Fig.~(\ref{fig:1}b).  Because we are grabbing only select elements of the full $A$-matrix, the columns of $\tilde A$ must be renormalized such that they once again sum to one.  That is to say given any two bit-strings $\texttt{row}, \texttt{col} \in \vec{p}_{\rm noisy}$, the reduced matrix element $\tilde{A}_{\texttt{row},\texttt{col}}$ is given by

\begin{equation} \label{eq:reduced}
\tilde{A}_{\texttt{row}, \texttt{col}} = 
    \left\{
        \begin{array}{cc}
            \cfrac{A_{\texttt{row},\texttt{col}}}{\sum_{\substack{ k \in \vec{p}_{\rm noisy} \\ d(k,\texttt{col})\leq D}} A_{k, \texttt{col}}} & \ \  \scriptstyle d(\texttt{row},\texttt{col}) \leq D \\
            \\
            0 & \ \ \scriptstyle d(\texttt{row},\texttt{col}) > D
        \end{array}
    \right.
\end{equation}

where $D$ is the desired Hamming distance \footnote{Keeping all elements is equivalent to setting $D$ equal to the number of measured qubits, whereas $D=0$ yields the identity matrix.}.  Performing a finite number of circuit executions, even when measurement errors are weak, may result in completely redistributing probability away from some small magnitude elements in $\vec{p}_{\rm ideal}$ such that those elements are not present in $\vec{p}_{\rm noisy}$; the solution vector will be missing these elements. However as we will show, for typical numbers of circuit samples this effect is minimal.

\section{Matrix-free solution}\label{sec:free}
Although $\tilde A$ is much smaller than the original, when sampling circuits with wide probability distributions many times, or executing on systems with appreciable error rates, it is possible that $\tilde A$ itself may become too costly to explicitly construct.  Fortunately, being able to grab elements of $A$ individually, Eq.~(\ref{eq:matelem}), allows us to take advantage of matrix-free iterative techniques \cite{saad:2003}.  The time to solution for iterative methods greatly depends on the properties of $\tilde A$.  $A$-matrices obtained from present day cloud-accessible platforms nominally have strict diagonal dominance $|A_{\tt row, \tt row}| > \sum_{\tt col \neq \tt row}|A_{\tt row, \tt col}|~\forall~\tt row$, and are readily solved by simple iterative methods such as Jacobi iteration \cite{saad:2003}.  However this condition does not hold in general for systems with large error rates.  Moreover, as the number of qubits grows so does the number of possible error channels (i.e. number of states at low Hamming distance) and it becomes harder to satisfy this stringent condition.  Thus general purpose methods such as generalized minimal residual (GMRES) \cite{gmres} or biconjugate gradient stabilized (BiCGSTAB) \cite{bicgstab} methods must be used.  Importantly these Krylov subspace methods require computing only the product $\tilde A\vec{p}_{\rm noisy}$, but not $\tilde A$ itself \cite{templates, saad:2003}.  However having strict diagonal dominance, or close to it, suggests that we can increase the rate of convergence by using a simple Jacobi preconditioner $P^{-1}$ to solve

\begin{equation}\label{eq:precond}
P^{-1}\tilde{A}\vec{x}=P^{-1}\vec{p}_{\rm noisy},
\end{equation}
where $P^{-1}$ is a diagonal matrix with $P^{-1}_{i,i}=1/\tilde{A}_{i,i}$ \cite{saad:2003}. In practice, Eq.~(\ref{eq:precond}) gives rapid convergence requiring only $\mathcal{O}(1)$ iterations for an absolute tolerance value of $10^{-5}$ while simultaneously dramatically reducing the memory requirements for mitigation.

\section{Uncertainty estimates}\label{sec:overhead}
Mitigating measurement errors does not come for free. Rather it results in an increase in the uncertainty of repeated measurement outcomes that must be compensated for by increasing the number of times the circuit is sampled.  This mitigation overhead $\mathcal{M}$ is determined by the one-norm of the inverse of the A-matrix $\mathcal{M}=||A^{-1}||^{2}_{1}$ \cite{bravyi:2020}, and gives an upper bound on the standard deviation of an observable $\sigma_{O}\leq \sqrt{\mathcal{M}/s}$, where $s$ is the number of samples.  Not wanting to construct $\tilde{A}^{-1}$, here we use the iterative Hager-Higham algorithm \cite{hager:1984, higham:1988} for estimating $||\tilde{A}^{-1}||_{1}$ using only linear systems of equations involving $\tilde{A}$ and $\tilde{A}^{T}$.  When using direct factorization, the LU decomposition of $\tilde{A}$ can be cached, and thus the overall run-time is $\sim 2x$ longer than mitigation alone.  However, for iterative methods, the overhead is between 4-10x longer depending on how many steps the Hager-Higham routine requires.  Although this method gives a lower-bound on $\sqrt{\mathcal{M}}$, in practice it is often exact or nearly so (see related discussion in Ref.~\cite{higham:1988}).  Because our truncation method selects only those rows and columns from $\vec{p}_{\rm noisy}$, the one-norm of $||\tilde{A}^{-1}||_{1}$, and thus mitigation overhead, is dependent on the circuit being executed and the noise properties of the device on which it is run.

\section{Demonstrations}\label{sec:demos}
Our method \cite{m3repo} is implemented with NumPy \cite{numpy}, SciPy \cite{scipy}, and Cython \cite{cython}, and makes use of Qiskit \cite{Qiskit} for calibration circuit construction and execution. All timing data is taken on a quad core Intel i3-10100 system with $32~\rm Gb$ of memory with NumPy and SciPy compiled against OpenBlas \cite{openblas}.

\begin{figure}[t]
\includegraphics[width=\columnwidth]{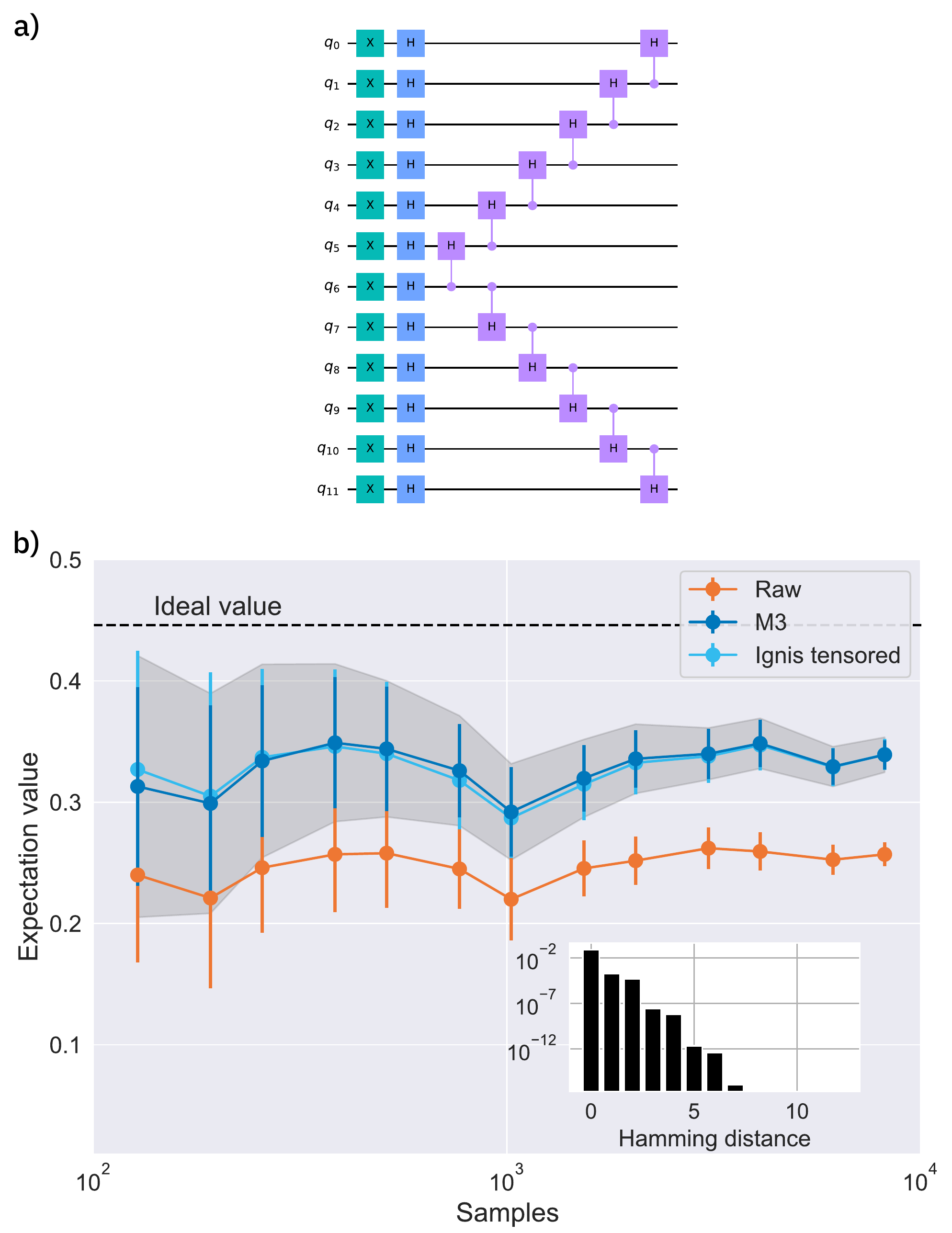}
\caption{a) 12-qubit circuit with an exact distribution of 43 unique bit-strings, and an expectation value of $\sim 0.446$. Measurements are omitted for brevity.  b) Expectation values before and after mitigating measurement errors for the circuit in (a) executed 100 times on the IBM Quantum Kolkata system while varying the number of samples per circuit.  Error bars show one standard deviation, while the shaded region gives error bounds determined by the computed mitigation overhead and number of samples.  The same calibration data is used for both Ignis and M3 and was computed using $8192$ samples per circuit.  Inset figure shows the absolute error when truncating $\tilde{A}^{(T)}$ to a given Hamming distance for the data collected at 8192 samples.}
\label{fig:2}
\end{figure}

We begin by showcasing the veracity of our method by comparing expectation values for the circuit shown in Fig.~(\ref{fig:2}a) computed with the full-space tensored method from Qiskit Ignis \cite{Qiskit} along with our method called M3 \footnote{This stands for: \textbf{M}atrix-free \textbf{M}easurement \textbf{M}itigation.} varying the number of samples taken per circuit.  Here the circuits are executed on the 27 qubit IBM Quantum Kolkata system, mapping the virtual circuit qubits to physical qubits $[1,4,7,10,12,13,14,11,8,5,3,2]$. The calibration data and raw input samples are identical for both mitigation methods. In Fig.~(\ref{fig:2}b) we see that, despite using \emph{at most} $371$ bit-strings ($9\%$ of the full-dimensionality), M3 closely matches the full-dimensional $A^{(T)}$ results over the entire range of samples, and agrees remarkably well with the Ignis results for the numbers of executions typically employed in practice, $>1000$.  An example $\tilde{A}^{(T)}$ generated by M3 is presented in App.~(\ref{app:Atilde}).  The sub-space reduction results in a run-time performance improvement as well, with Qiskit Ignis mitigation taking $\sim 3~\rm s$ per circuit, where as M3 took at most $\sim 7~\rm ms$.  Additionally, we see that the uncertainty bound given by $\mathcal{M}$ closely matches the experimental values, and verifies the use of this technique in reporting faithful error bounds.  The inset of Fig.~(\ref{fig:2}b) also shows that, for a tolerance of $\leq 10^{-5}$, $\tilde{A}^{(T)}$ is well-approximated by keeping only terms out to $D=3$.

\begin{figure}[t]
\includegraphics[width=\columnwidth]{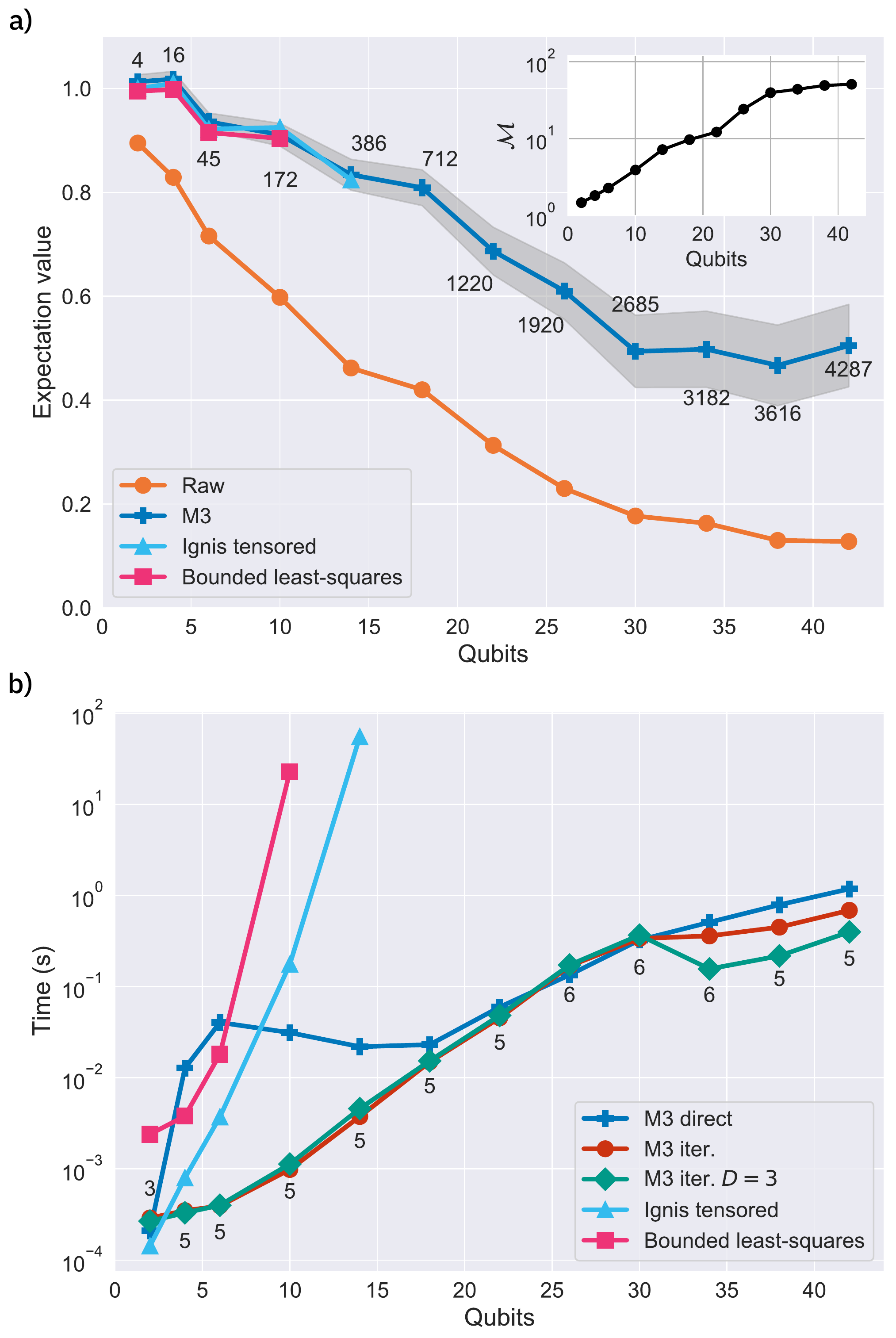}
\caption{a) Comparison of expectation values for GHZ states out to 42 qubits on the IBM Quantum Brooklyn system using M3, and Qiskit Ignis tensored and bounded least-squares methods.  M3 direct, iterative, and a $D=3$ Hamming approximation all yield the same values to a tolerance of $\sim 10^{-5}$ and the associated shaded region shows the error bound from $\mathcal{M}$.  Numbers alongside the M3 data show the number of bit-strings in $\vec{p}_{\rm noisy}$ for each number of qubits.  Inset shows the computed mitigation overhead $\mathcal{M}$.  All calibration data and GHZ circuit execution was performed using $8192$ shots per-circuit. b) Timing information (best of three runs) for the mitigation methods presented in (a).  Annotated numerical values indicate the number of iterations needed for a tolerance $\leq 10^{-5}$.  The full and $D=3$ iterative solutions used the same number of iterations in all cases.}
\label{fig:3}
\end{figure}

We now demonstrate the scalability of our M3 method by mitigating GHZ states out to 42 qubits on the 65 qubit IBM Quantum Brooklyn system.  Details of this experiment are in App.~(\ref{app:ghz}).  In Fig.~(\ref{fig:3}) we compare M3 along with the Qiskit Ignis tensored and bounded least-squares methods \footnote{We have modified the Qiskit least-squares method to use $\vec{p}_{\rm noisy}$ as the starting vector as opposed to a random vector.  This gives a 3x or more improvement in runtime.}.  Only M3 allows for mitigating errors beyond 14 qubits due to algorithmic breakdown or extreme run-times, for the tensored and least-squares methods respectively, and shows the importance of performing measurement mitigation for large-scale experiments.  The overall expectation values drop as the circuit depth increases where gate errors and decoherence, effects measurement mitigation cannot resolve, start to dominate.

The mitigation overhead, inset of Fig.~(\ref{fig:3}a), shows exponential scaling at small numbers of qubits after which the overhead begins to plateau. The exponential scaling arises as the diagonal elements of $\tilde{A}$, formed from the product of $N$ probabilities, Eq.~(\ref{eq:matelem}), are effectively being inverted when computing $\tilde{A}^{-1}$, with additional contributions coming from elements close in Hamming distance.  Provided that the bit-strings in $\vec{p}_{\rm noisy}$ sample sufficient portions of these short Hamming distance elements, the re-normalization used in obtaining $\tilde{A}$ is small, and one recovers the exponential scaling shown for the full-dimensional A-matrix \cite{bravyi:2020}.  However if this is not the case then re-normalization increases the magnitude of the elements in $\tilde{A}$ (in particular the diagonal elements) suppressing the exponential growth in $\mathcal{M}$.

Figure~(\ref{fig:3}b) details the timing across the different mitigation methods.  We see that M3 greatly improves the computed expectation values while taking at most $1.2$ seconds to compute at $42$ qubits.  When computing the mitigation overhead, the total time increased to $2.4$ and $4.5$ seconds for the direct and iterative solutions at $42$ qubits, respectively (not shown); an extraordinary improvement upon the exponential run times observed for the Qiskit methods.  As with the example in Fig.~(\ref{fig:2}), a $D=3$ Hamming approximation well captures the full mitigation process to the desired tolerance, and performs best at large numbers of bit-strings where the overhead from computing the Hamming distance between elements starts to become less than the cost of additional floating-point multiplications.  For the M3 results, the rate at which the run times increase begins to slow down at larger numbers of qubits following a similar slow down in the number of unique bit-strings observed, Fig.~(\ref{fig:3}a), when additional qubits are added.

Finally, we note that at $42$ qubits, storing a full $2^{N}$-vector of single-precision floating point values for $\vec{p}_{\rm noisy}$ requires $16~\rm TiB$ of memory; well beyond the limits of our computer on which the mitigation is implemented, but is amenable to storage on a supercomputer.  Juxtapose that with storing a sparse representation of $A$ using, for example, compressed sparse column (CSC) format out to $D=3$. This requires $\sim 580~\rm PiB$ of memory, which is $120$x more than that available in the Fugaku supercomputer \cite{fugaku} [see App.~(\ref{app:memory}) for details].  In contrast, the M3 iterative method uses $\sim 1~\rm MiB$ of storage, highlighting the benefit of the techniques presented here for mitigating measurements at scales amenable to demonstrations of quantum advantage.

\begin{acknowledgments}
We thank Doug McClure, David McKay, and Matthew Treinish for helpful discussions.  Figures~(\ref{fig:2}-\ref{fig:corr}) are produced using Matplotlib \cite{matplotlib}.
\end{acknowledgments}

\appendix

\section{Example A-matrices}\label{app:amat}

\subsection{Complete A-matrix}\label{app:fullA}

\begin{figure}[h]
\includegraphics[width=\columnwidth]{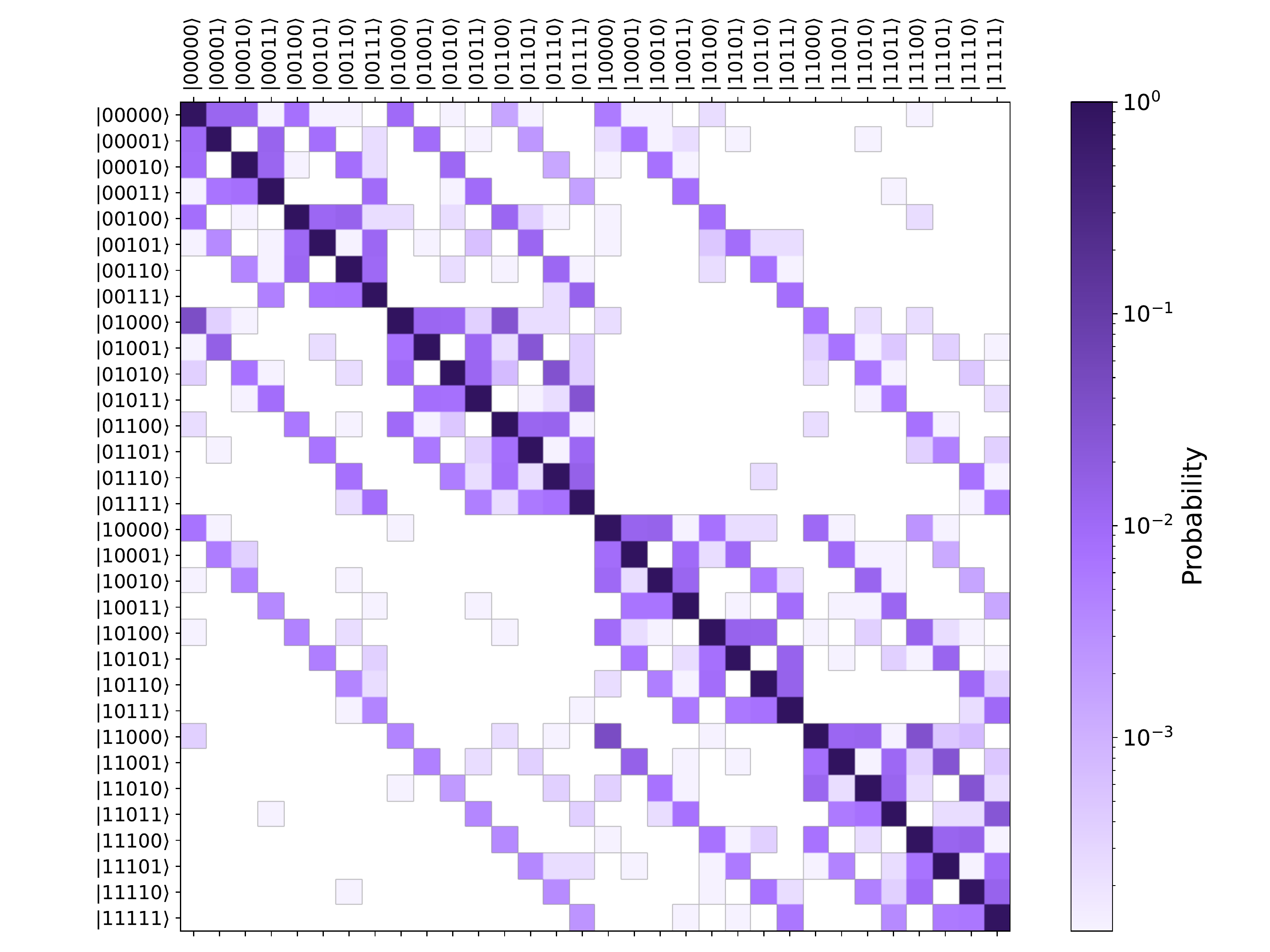}
\caption{Full A-matrix for qubits $0\rightarrow 5$ on the IBM Quantum Kolkata system.  A circuit for each of the 32 bit-strings was executed 8192 times to fill in the columns.}
\label{fig:supp1}
\end{figure}

An example complete A-matrix computed by running $2^{N}$ circuits, one for each computational basis state, on the IBM Quantum Kolkata system is given in Fig.~(\ref{fig:supp1}).  Because of finite sampling, the matrix is nominally sparse, and only those elements close in Hamming distance have appreciable transition probabilities.  The matrix has strict diagonal dominance and thus is guaranteed to be invertible.

\subsection{Tensored A-matrix}\label{app:tensorA}

The A-matrix corresponding to tensored measurement errors, $A^{(T)}$ is constructed by taking the tensor product of single-qubit calibration matrices given by Eq.~(\ref{eq:1qcal}) in the main text.  Unlike the complete A-matrix, $A^{(T)}$ contains only non-zero elements unless one or more qubits has no reported measurement error for $P^{(k)}_{0,1}$ and/or $P^{(k)}_{1,0}$; $A^{(T)}$ expects every element of $\vec{p}_{\rm noisy}$ to have a non-zero entry.  Like Fig.~(\ref{fig:supp1}), the matrix in Fig.~(\ref{fig:supp2}) is strictly diagonally dominant, and indicates that transitions between elements close in Hamming distance are more likely.  Data for Fig.~(\ref{fig:supp2}) was taken immediately after that shown in Fig.~(\ref{fig:supp1}).

\begin{figure}[t]\label{fig:supp2}
\includegraphics[width=\columnwidth]{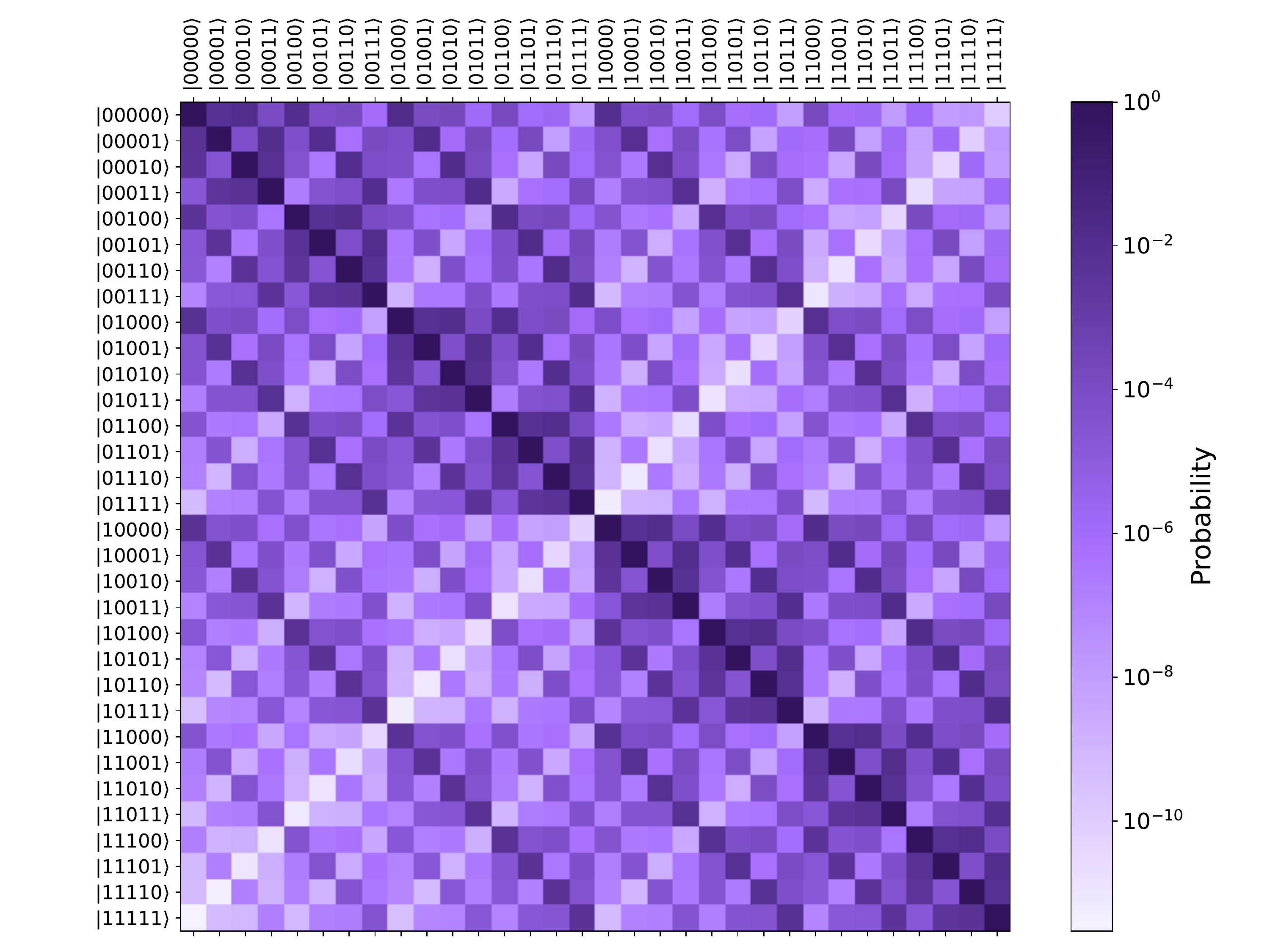}
\caption{Tensored A-matrix $A^{(T)}$ for qubits $0\rightarrow 5$ on IBM Quantum Kolkata.}
\label{fig:supp2}
\end{figure}

\subsection{Example 12 qubit truncated A-matrix}\label{app:Atilde}

\begin{figure}[t]
\includegraphics[width=\columnwidth]{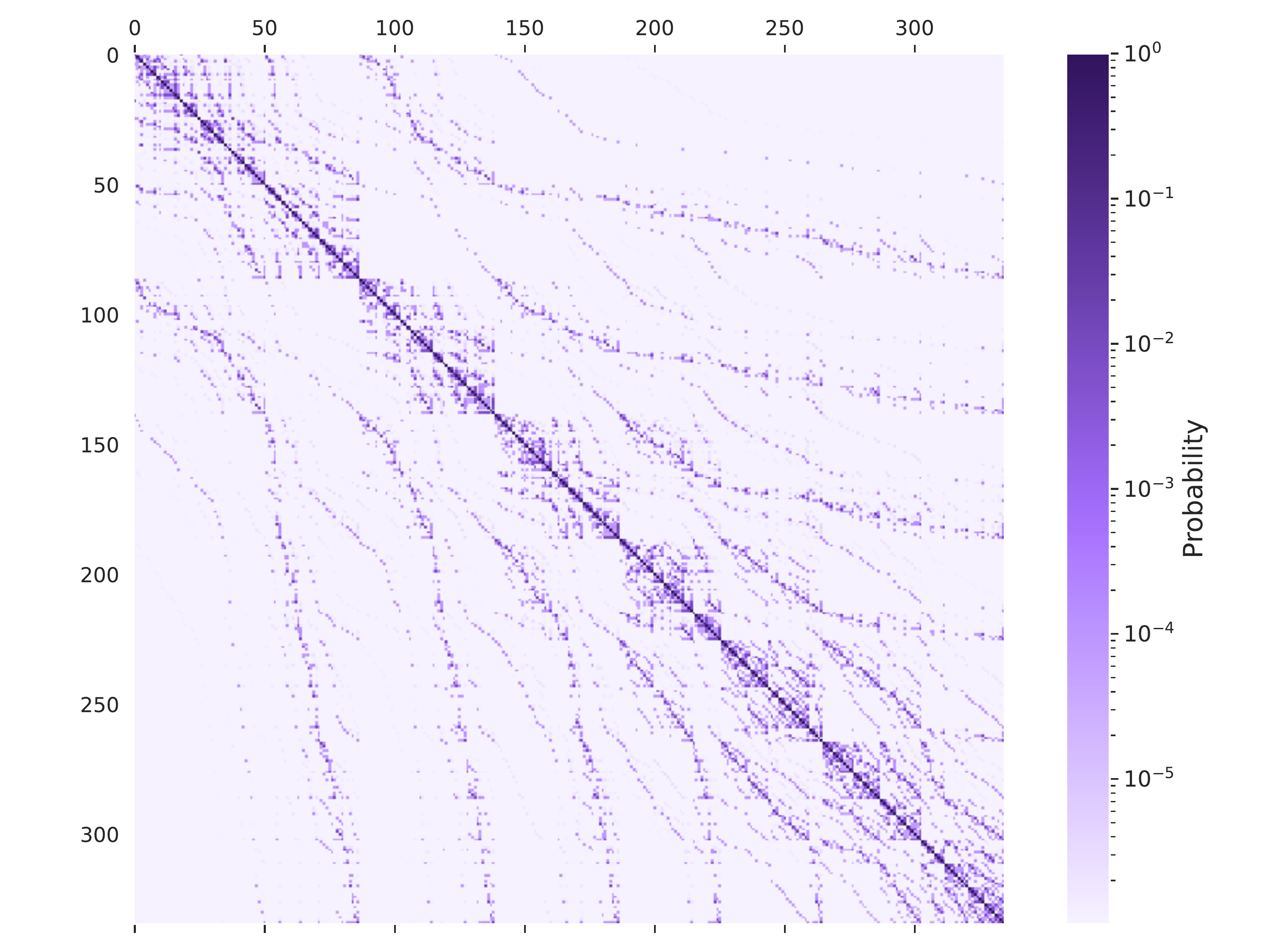}
\caption{Sample $\tilde{A}$ from the M3 results in Fig.~(\ref{fig:2}b) taken at 8192 samples.  The matrix contains $341$ bit-strings out of a possible $4096$.  Rows and columns are sorted in lexicographical order in terms of their bit-string representation.}
\label{fig:app3}
\end{figure}

Figure~(\ref{fig:app3}) shows one of the $100$ truncated $\tilde{A}^{(T)}$ used in the M3 mitigation performed in Fig.~(\ref{fig:2}) at $8192$ counts.  This matrix is also strictly diagonally dominant.  For each circuit execution, the number of elements in $\tilde{A}^{(T)}$ may vary, as can their associated amplitudes.

\section{Correlated errors}\label{app:correlated}
Although we have focused on tensored mitigation techniques, our method is equally capable of handling correlated errors provided that matrix elements $\tilde{A}_{\tt row, \tt col}$ can be obtained by their bit-string values as done in Eq.~(\ref{eq:matelem}).  IBM Quantum systems operating normally are dominated by uncorrelated measurement error, and are thus well mitigated by the tensored A-matrix methods presented in the main text.  It is possible however to purposely induce correlated errors into the readout process and explore correlated mitigation strategies.

To take into account pairwise correlated errors, we can modify Eq.~(\ref{eq:matelem}), grabbing elements using

\begin{equation}\label{eq:correlated}
A_{\tt row, \tt col}^{(C)} = \frac{1}{\binom{N}{2}}\sum_{k=1}^{N-1}\sum_{l=0}^{k-1}C_{kl}[a^{kl}, b^{kl}]\prod^{N-1}_{\substack{m=0 \\ m\neq k, l}}S_m[q_m, q'_m],
\end{equation}
where $\texttt{row}=q_{N-1}q_{N-2}..q_0$ and $\texttt{col}=q'_{N-1}q'_{N-2}..q'_0$. Here, the $S_m$ are defined as in Eq.~(\ref{eq:1qcal}), and $C_{kl}$ are a $4x4$ stochastic matrix (local noise matrix) between qubits $k$ and $l$ where $a^{kl}=2q_k+q_l$,~ $b^{kl}=2q'_k+q'_l$, respectively. The elements of $C_{kl}$ are obtained following Section $V$ of Ref.~\cite{bravyi:2020}.

\begin{figure}[t]
\includegraphics[width=\columnwidth]{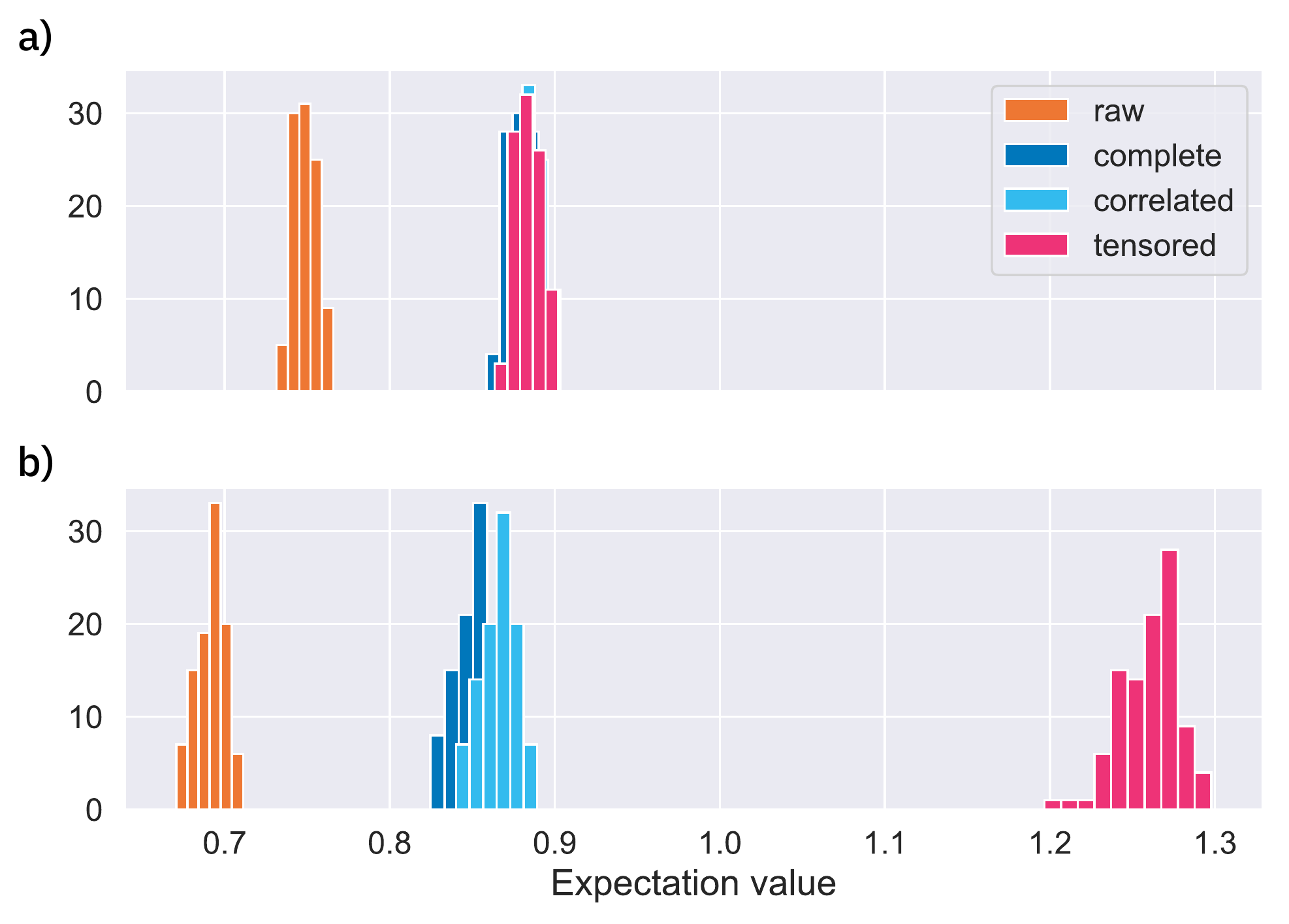}
\caption{a) Histogram of expectation values collected from executing an 8 qubit GHZ circuit 100 times on the IBM Quantum F5.11 Kolkata system under normal operating conditions and using qubits $[8, 5, 3, 2, 1, 4, 7, 6]$ and $8192$ samples per circuit.  Raw results are mitigated with the complete ($A$), correlated M3 ($\tilde{A}^{(C)}$), and tensored M3 ($\tilde{A}^{(T)}$). b) Repeated experimental results in the presence of correlated readout errors generated by degraded readout.}
\label{fig:corr}
\end{figure}

IBM Quantum Kolkata, a Falcon 5.11 series system, has readout output multiplexing ratios ranging from  $3:1$ to $5:1$, with readout frequencies in a shared output typically separated by $50$-$60~\rm MHz$. This separation is much larger than the average cavity linewidth $\kappa$ ($5.6~\rm MHz$) and dispersive shift $\chi$ ($1.6~\rm MHz$), which when combined enable short $330~\rm ns$ readout pulses for all qubits. Using default readout pulse amplitudes, calibrated to optimize fidelity while maintaining QND readout,  Fig.~(\ref{fig:corr}a) shows expectation values produced by executing an 8 qubit GHZ circuit 100 times using qubits $[8, 5, 3, 2, 1, 4, 7, 6]$ that span two multiplex readout groupings. Importantly we perform mitigation using the complete $A$-matrix, $\tilde{A}^{(C)}$, as well as $\tilde{A}^{(T)}$, with results in agreement with uncorrelated errors largely dominating the readout process. 

Intentionally increasing readout pulse amplitudes by approximately 2x from the optimized values results in correlated readout errors. 
With these larger readout amplitudes, while there is no appreciable change in average readout fidelity ($< 0.2\%$) compared to the default setting, there is a substantial uptick in non-QNDness.  In Fig.~{(\ref{fig:corr}b)} we re-run our 8 qubit GHZ experiment from Fig.~{(\ref{fig:corr}a)} under these new conditions.  While the correlated M3 method using Eq.~(\ref{eq:correlated}) well captures the the correlated readout errors, as evident by agreement with the complete A-matrix, the tensored mitigator strongly over-corrects due to the misalignment of probabilities in $\tilde{A}^{(T)}$ with those actually present in the system.  Although Eq.~(\ref{eq:correlated}) works for both uncorrelated and pairwise correlated errors, each matrix element requires $\mathcal{O}\left(N^{2}\right)$ floating-point evaluations, as opposed to $N$ in Eq.~(\ref{eq:matelem}).

\section{42 qubit GHZ demonstration}\label{app:ghz}

\subsection{Experimental details}
\begin{figure}[b]
\includegraphics[width=\columnwidth]{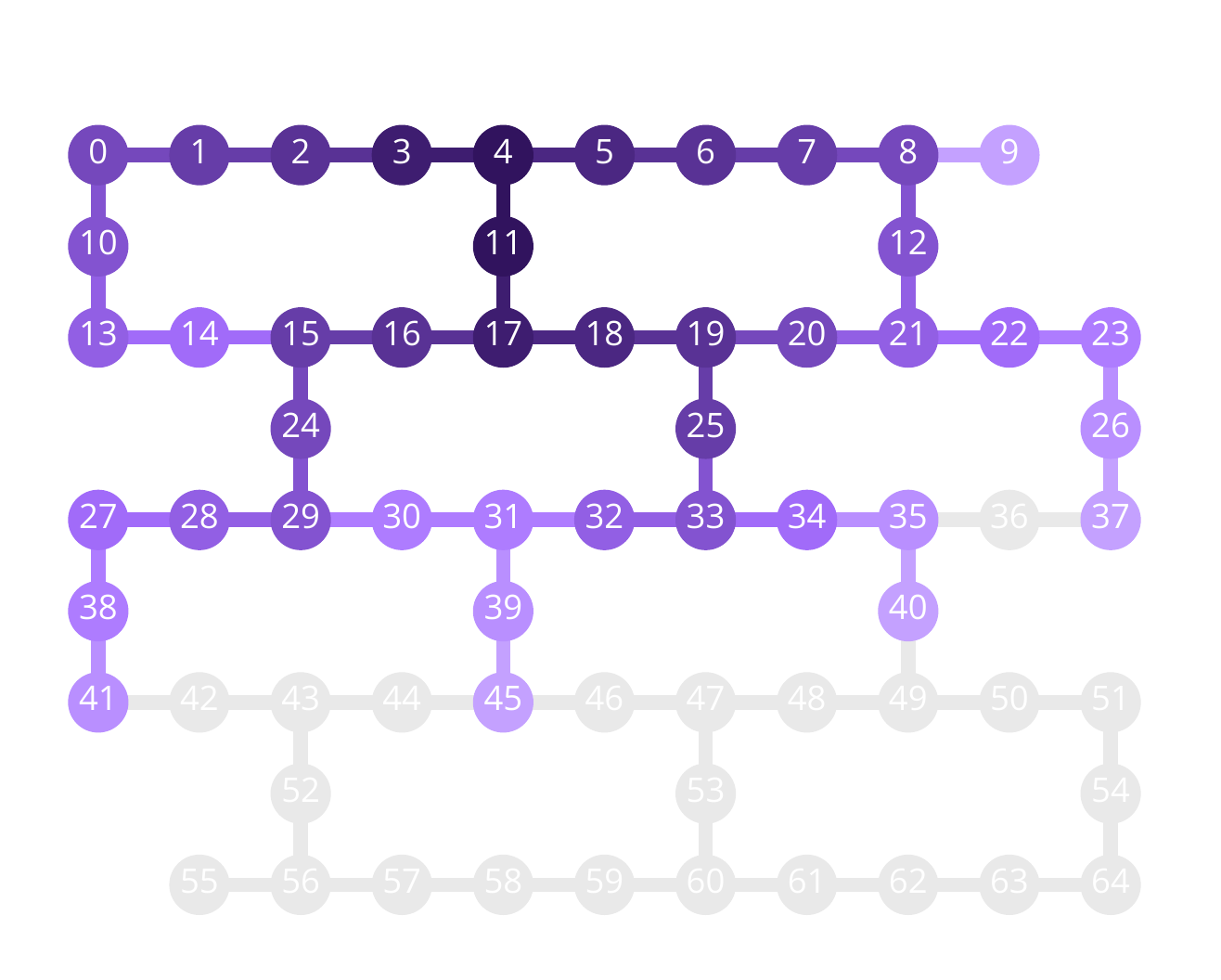}
\caption{Qubits used in generating GHZ states on the IBM Quantum Brooklyn system.  Gates entangled earlier are color coded darker.  The initial two qubit GHZ state is between qubits 11 and 4, while the final 42 qubit GHZ state is formed by adding qubits 9, 37, 40, and 45 to the previous iteration.}
\label{fig:supp4}
\end{figure}

Experiments are run on the 65 qubit IBM Quantum Brooklyn system.  GHZ states were prepared starting with a Hadamard gate on qubit $11$ and entangling additional qubits as shown in Fig.~(\ref{fig:supp4}).  After entangling the first 6 qubits, this pattern allows for increasing the GHZ state by 4 qubits per layer.  The average assignment and CNOT error rates across the qubits used is $2.15\%$ and $1.01\%$, respectively.

\subsection{Memory requirements for storing full 42 qubit A-matrix to $D=3$}\label{app:memory}
Including elements up to a Hamming distance of three requires 

\begin{equation}
\binom{42}{0} + \binom{42}{1} + \binom{42}{2}+ \binom{42}{3} = 12384
\end{equation}

elements in each of the $2^{42}$ columns.  Storing these values using single-precision floating-point numbers, 4-bytes per entry, requires $193.5~\rm PiB$ of memory.  In addition, we must also specify the row and column indices for these values.  In compressed sparse column (CSC) format we need $2^{42}+1$ elements, the difference of which specify the number of non-zero elements in each of the columns.  Lastly, we also need the row index for each non-zero matrix element.  At 42 qubits, the size of these indices cannot be stored using 32-bit integers, and we must use 64-bits per entry.  Storing these values requires an additional $387~\rm PiB$ of memory. The total memory required is therefore $580.5~\rm PiB$ and is $\sim120$x larger than the $4.85~\rm PiB$ of memory on the Fugaku supercomputer \cite{fugaku}.

\bibliography{refs}
\end{document}